\documentclass[aip,jcp,reprint,amsmath,amssymb,floatfix]{revtex4-2}
\usepackage{graphicx,bm,mathtools}
\usepackage[dvipsnames]{xcolor}
\usepackage[colorlinks=true,allcolors=blue]{hyperref}
\usepackage{bbm}
\usepackage{microtype}
\usepackage{comment}
\usepackage{dcolumn} 
\usepackage{booktabs}
\usepackage{multirow}

\newcommand{\Ylm}{Y_{lm}}
\newcommand{\Ylmc}{Y^{*}_{lm}}
\newcommand{\dO}{\,\mathrm{d}\Omega}
\newcommand{\ii}{\mathrm{i}}
\newcommand{\ee}{\mathrm{e}}
\newcommand{\kB}{k_{\rm B}}
\newcommand{\eps}{\varepsilon}
\newcommand{\epsw}{\eps_{w}}
\newcommand{\epsp}{\eps_{p}}
\newcommand{\lB}{\ell_{B}}
\newcommand{\Om}{\Omega}

\newcommand{\sig}{\sigma}
\newcommand{\Gmat}{\mathsf{G}}
\newcommand{\Kmat}{\mathsf{K}}
\newcommand{\Imat}{\mathbbm{1}}
\newcommand{\svec}{\mathbf{s}}
\newcommand{\szero}{\mathbf{s}^{(0)}}
\newcommand{\sstar}{\mathbf{s}_{\star}}
\newcommand{\Cnull}{C_{0}}
\newcommand{\Cd}{C_{d}}

\newcommand{\il}{i_{l}}

\newcommand{\threej}[6]{\begin{pmatrix} #1 & #2 & #3 \\ #4 & #5 & #6 \end{pmatrix}}

\DeclareMathOperator{\diag}{diag}

\newcommand{\pKa}{\mathrm{p}K_{a}}
\newcommand{\Bt}{B_{22}}

\newcommand{\rev}[1]{\textcolor{black}{#1}}

\begin{document}

\title{A closed-form theory of charge-regulated electrostatic interactions
between anisotropically charged spheres}

\author{Anže Božič}
\email{anze.bozic@ijs.si}
\affiliation{Department of Theoretical Physics, Jo\v{z}ef Stefan Institute, Ljubljana, Slovenia}

\date{\today}

\begin{abstract}
Proteins and many other colloids carry ionizable surface groups that are both spatially inhomogeneous (patchy) and pH-responsive (charge-regulating). In the description of the electrostatic interactions between such particles, these two aspects are often treated separately, especially from a theoretical perspective. We present a unified, closed-form theory that unites charge regulation and patchiness within the linearized Poisson--Boltzmann approximation. For spherical particles with anisotropic distributions of titratable sites, we derive both the leading-order mean-field interaction energy as well as the Kirkwood--Shumaker fluctuation interaction.  The theory exposes a qualitatively new effect between two electroneutral anisotropic particles: proximity-induced charge regulation generates a net monopole, and splits orientational branches that are {mirror-symmetric} in any linear fixed-charge model. The accompanying correction to the interaction energy has no fixed sign---it weakens like-charge repulsion but can deepen an existing attraction or create one where fixed charge predicts none. We benchmark three components of the theory against published results, as no existing simulation probes their combination, which would capture the central prediction of the theory. Lastly, we apply the theory to two proteins---lysozyme and $\alpha$-chymotrypsinogen A---and show that it not only reproduces the general trends of the measured second virial coefficients but also predicts how charge regulation expands the range of pH and screening strength where the protein--protein interaction is attractive.
\end{abstract}

\maketitle

\section{Introduction}
The strength and sign of the electrostatic interaction between two protein or colloidal particles govern their behavior in solution~\cite{Kimura_2019,lebdioua2021jcis,naderi2020self,li2022modulating,guo2021quantifying,zhang2020assembly,zhou2018electrostatic,kim2024surface}, which is routinely summarized and characterized by the osmotic second virial coefficient $\Bt$~\cite{george1994,neal1999}. The classical description of this interaction, descending from the Debye--H\"uckel (DH) theory of screened electrostatics~\cite{debye1923} through DLVO theory~\cite{derjaguin1941,verwey1948}, treats the particles as uniformly charged spheres. This neglects two important features of real macromolecular surfaces. First, the charge is spatially inhomogeneous: ionizable groups cluster into patches, producing a surface whose low-order multipole moments can dominate the interaction even when the net charge is small---a feature central to the design of patchy colloids~\cite{glotzer2007,hong2006,bianchi2011,vissers2013predicting,Cruz_2016,abrikosov2017steering,Brunk_2020} and increasingly implicated in the aggregation and developability of therapeutic proteins~\cite{ausserwoger2023}. Crucially, environmental factors like pH alter not just the magnitudes of these single-particle multipole moments but also the spatial orientations of their principal axes~\cite{bozic2017ph}; even specific ion binding can render the interactions of an otherwise uniformly charged protein anisotropic~\cite{lund2016anisotropic}. Second, the charge is responsive: each group titrates according to its intrinsic dissociation constant $\pKa$ and the local electrostatic potential, so that the charges of two approaching particles {regulate}---a mechanism which has been developed extensively for interacting surfaces~\cite{linderstromlang1924,kirkwood1952,ninham1971,chan1975,healy1980,carnie1993,behrens1999,lund2005,lund2013} and confirmed by direct force measurements~\cite{popa2010,trefalt2016}.
 
Theoretical descriptions of macromolecular electrostatics, however, rarely include both features simultaneously. The charge regulation (CR) mechanism is most often formulated for homogeneous surfaces, and various theoretical approaches all describe how a uniform surface or a globally titrating protein responds to its neighbor, but carry no information about where on the surface the responsive groups sit~\cite{chan1975,healy1980,carnie1993,pericet2004,carnie1994,behrens1999,lund2005,lund2013,markovich2016,podgornik2018,smallenburg2011jcp,bakhshandeh2019}. The principal exception is the work of Boon and van Roij~\cite{boon2011,boon2012electrostatics}, who showed in a planar geometry that for two interacting surfaces, a chargeable patch induces charge on its neighbor strongly enough to turn long-range repulsion into short-range attraction. Anisotropic electrostatics, conversely, is almost always formulated at fixed charge: different descriptions of charge patchiness resolve the multipole structure of the surface charge density but hold every multipole fixed as two particles approach each other~\cite{phillies1974,mcclurg1998,hoffmann2004molphys,boon2010jpcm,bianchi2011sm,degraaf2012,bozic2013,hoppe2013,bianchi:2015,hieronimus2016jcp,everts2020,obolensky2021,siryk2021,mathews2022molsim,zhou2023interaction,popov2023jpcb,gnidovec2025}. This is often true also of simulations using charged patchy particle models against which such theories are benchmarked~\cite{yigit2015,blanco2016}. Methods that capture both features at once---numerical Poisson--Boltzmann (PB) calculations on structure-based protein models~\cite{neal1998,neal1999} and constant-pH or grand-reaction Monte Carlo (MC) simulations~\cite{pineda2025}---are faithful but computationally heavy, reporting the combined effect at a handful of state points and leaving the mechanism to be inferred from the output. Many-body simulations have established that the combination of CR and patchiness matters~\cite{curk2021,yuan2022} and in adsorption geometries CR and charge-patch effects have been found to act jointly---and non-additively---near the isoelectric point~\cite{lunkad2022}. Experimentally, static light scattering (SLS) on lactoferrin has been used to trace a non-monotonic salt dependence of $\Bt$ to a small, highly complementary charged patch competing with net-charge repulsion~\cite{li2015}.

Here, we develop a closed-form theoretical description of the pair interaction energy between two spherical particles that unifies CR with anisotropic distribution of surface titratable sites at the level of the linearized PB (DH) description. The same framework also delivers the Kirkwood--Shumaker (KS) charge-fluctuation interaction~\cite{kirkwood1952,adzic2014,adzic2015} of anisotropic, multi-species surfaces in closed form. Two consequences stemming from the theory have no counterpart in the linear fixed-charge description: two electroneutral anisotropic particles generally acquire a net monopole as they approach each other, and this coupling splits orientational branches that are mirror-symmetric at fixed charge. The paper is organized as follows: Section~\ref{sec:theory} develops the unified framework. Section~\ref{sec:results} presents the results in three stages---minimal dipolar and quadrupolar models that isolate the regulation-induced monopole, its symmetries, and the sign of the CR correction; three benchmarks based on published works; and the structure-based application to proteins. Section~\ref{sec:discussion} predominantly discusses the scope and limitations of the linear closure relation.

\section{Unified theoretical framework}
\label{sec:theory}
\subsection{System setup}

\begin{figure}[tb]
\centering
\includegraphics{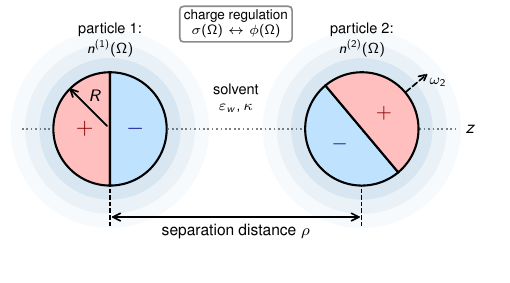}
\caption{Sketch of the system. Two anisotropically charged, charge-regulating spherical particles of radius $R$ are positioned along the $z$ axis at a center-to-center distance $\rho$. Dipole symmetry (Janus particles) is chosen here for clarity; the sketch does not represent the actual (continuous) surface charge distribution. The particles are embedded in a solvent (water) with dielectric constant $\varepsilon_w$ and monovalent $1:1$ salt concentration $c_0$, the latter giving rise to the inverse Debye screening length $\kappa$. Each particle ($i=1,\,2$) carries a surface number density of a set of different ionizable species $j$, $n_j^{(i)}(\Omega)$, which together give rise to a surface charge density on each particle $\sigma^{(i)}(\Omega)$. The particles can be rotated by $\omega_i$ from their initial states; only $\omega_2$ is shown for clarity. CR couples the site densities of ionizable species---and thus the surface charge---with the local (dimensionless) surface electrostatic potential $\phi(\Omega)=\beta e\,\psi(\Omega)$.
}
\label{fig:sketch}
\end{figure}

Two anisotropically charged spherical particles of radius $R$ are in a monovalent $1$:$1$ electrolyte with bulk concentration $c_0$ (inverse Debye length $\kappa=\sqrt{8\pi\ell_BN_Ac_0}$) and exterior permittivity $\epsw$ (Fig.~\ref{fig:sketch}). Here, $N_A$ is the Avogadro number, $\lB=\beta e^{2}/4\pi\epsw\varepsilon_{0}$ is the Bjerrum length, with $\beta=1/k_\mathrm{B}T$, $k_\mathrm{B}$ the Boltzmann constant and $T$ temperature; we set $\varepsilon_w=78$ for water and $T=298$~K as the room temperature. For convenience, the particles are aligned along the $z$ axis at a distance $\rho$ between their centers of mass.

Every field $f$ on the surface of the particles is expanded in spherical harmonics \rev{$Y_{lm}$}, $f(\Om)=\sum_{lm}f_{lm}\Ylm(\Om)$, with the expansion coefficients given by $f_{lm}=\oint f(\Omega)Y^*_{lm}(\Omega)\,\mathrm{d}\Omega$; $\Omega$ is the solid angle. We introduce $\phi(\Om)\equiv\beta e\,\psi(\Om)$ as the dimensionless surface potential and $\sigma^{(i)}(\Omega)$, $i=1,\,2$, as the surface charge distributions on the two particles. We collect the multipole expansion of the two surface charge densities in a stacked \rev{charge} vector
\begin{equation}
    \svec = \big(\{\sigma^{(1)}_{lm}\};\, \{\sigma^{(2)}_{lm}\}\big),
\qquad l \geqslant 0,\ |m| \leqslant l;
\label{eq:svec}
\end{equation}
we also have $\sigma^{(i)}_{lm}=(-1)^m\sigma^{(i)\,*}_{l,-m}$ due to the reality of $\sigma^{(i)}(\Omega)$. Each of the two surface charge distributions can be arbitrarily (and independently) rotated with respect to the initial state by a Wigner--$D$ matrix,
\begin{eqnarray}
\sig_{lm}^{(i)}\ &\rightarrow & \sum_{m'}D^{\,l}_{m'm}(\alpha_i,\beta_i,\gamma_i)\,\sig_{lm'}^{(i)},
\\
D^{\,l}_{m'm}(\alpha_i,\beta_i,\gamma_i)&=&\ee^{-\ii m'\alpha_i}\,d^{\,l}_{m'm}(\beta_i)\,\ee^{-\ii m\gamma_i},
\label{eq:wignerD}
\end{eqnarray}
with Euler angles $\omega_{i}=(\alpha_{i},\beta_{i},\gamma_{i})$ in the $zyz$ convention~\cite{varshalovich1988}.

The derivation of the unified framework describing charge-regulated electrostatic interactions between anisotropically charged spheres proceeds in four steps: {\em (i)}
write the DH interaction energy of the pair as one quadratic form in $\svec$; {\em (ii)} promote the fixed surface charge to a local, linear, potential-dependent constitutive law; {\em (iii)} combine and minimize; and {\em (iv)} integrate out the charge fluctuations. Parts of the first two steps build on our previous work~\cite{bozic2013,bozic2018,bozic2021} and are recreated here for the sake of completeness, cast in a form that enables the derivation of the final result. \rev{Table~\ref{tab:quantities} contains an overview of key quantities used throughout the manuscript.}

\renewcommand{\arraystretch}{1.3} 
\begin{table}[!htp]
\caption{\label{tab:quantities}\rev{Overview of key quantities.}}
\begin{ruledtabular}
\begin{tabular}{lc}
symbol  & meaning \\
\midrule
$\sigma^{(i)}(\Omega)$ & surface charge density  of particle $i$ \\
\multirow{2}{*}{$\svec$} & stacked charge vector of multipole\\
&  coefficients of both particles \\
$\phi(\Omega)$ & dimensionless surface potential \\
$\sigma_0^{(i)}$, $\szero$ & bare charge (at zero surface potential) \\
$\Gmat$ & DH response operator (permeable) \\
$\Gmat^\mathrm{d}$ & dielectric DH response operator (impermeable) \\
$\mathcal{C}(\Omega)$ & capacitance density \\
$\Kmat$ & capacitance operator \\
$\svec_\mathrm{iso}$ & charge self-regulated in own potential \\
$\sstar$ & charge regulated in presence of interaction partner \\
$\Gmat_\star$ & DH$+$CR response operator \\
$V_\mathrm{int}$, $V_\mathrm{KS}$ & mean-field and KS interaction free energies \\
$Q_\mathrm{induced}$ & net charge induced due to interaction partner \\
$a_l$ & anisotropy contrast \\
\end{tabular}
\end{ruledtabular}
\end{table}

\subsection{Fixed-charge electrostatic interaction energy}
\label{ssec:DH}

Linearized solutions of the anisotropic two-sphere problem  have a long history~\cite{phillies1974,mcclurg1998,bozic2013,hoffmann2004molphys,degraaf2012}. We write here the known DH interaction energy $F_\mathrm{ES}$ of two arbitrarily charged spherical shells whose interior is filled with solvent (i.e., $\varepsilon=\varepsilon_w$ throughout), carrying a \emph{fixed} anisotropic surface charge distribution $\svec$~\cite{bozic2013}:
\begin{equation}
F_{\rm ES}=\frac{R^{2}}{2}\,\svec^{\dagger}\Gmat(\rho,\omega_{1},\omega_{2})\,\svec,
\label{eq:FES}
\end{equation}
where $\svec^\dagger$ denotes the conjugate transpose (with entries $\sigma_{lm}^{(i)\,*}$); matrix multiplication is implied throughout. The \rev{DH response} operator $\Gmat$,
\begin{equation}
\Gmat=\begin{pmatrix}\diag(g_{l}) & \Gmat_{12}\\[2pt]\Gmat_{21}&\diag(g_{l})\end{pmatrix},
\label{eq:Gblock}
\end{equation}
carries all of the separation ($\rho$) and orientation ($\omega_{1,2}$) dependence
through modified spherical Bessel functions and Wigner rotations; $\bm{\phi}=\beta e\,\Gmat\svec$ is the surface potential each $(l,m)$ mode feels. The diagonal (self) blocks $\Gmat_{ii}$ reduce to the single-particle mode response functions
\begin{equation}
g_{l}=\frac{\Cnull(l,\kappa R)}{\kappa\epsw\eps_{0}},
\label{eq:gl}
\end{equation}
where
\begin{equation}
\Cnull(l,x)=x^{2}\,i_{l}(x)\,k_{l}(x),
\label{eq:cnull}
\end{equation}
and $i_l$ and $k_l$ are the modified spherical Bessel functions in the standard convention~\cite{arfken2011mathematical}. The off-diagonal blocks of $\Gmat$ encode the $(l,m)$ mode of the potential induced at particle~$2$ by the mode $(p,m)$ of the charge of particle~$1$ (the azimuthal number is conserved because $\bm{\rho}\parallel\hat z$)~\cite{bozic2013}:
\begin{eqnarray}
\nonumber\big[\Gmat_{21}\big]_{(lm),(pm)}
&=&\frac{(-1)^{l+m}}{\kappa\epsw\eps_{0}}\sqrt{(2l+1)(2p+1)} \\
\nonumber&\times&\sum_{s}\Bigg[(2s+1)\,
\il(\kappa R)\,\frac{\Cnull(p,\kappa R)}{k_{p}(\kappa R)}\\
&\times& k_{s}(\kappa\rho)\,
\threej{l}{s}{p}{0}{0}{0}\threej{l}{s}{p}{m}{0}{-m}\Bigg],
\label{eq:G21}
\end{eqnarray}
where $|l-p|\leqslant s\leqslant l+p$ and $l+p+s=\mathrm{even}$; it also follows that $\Gmat_{12}=\Gmat_{21}^{\dagger}$.

\subsubsection{Extension to impermeable dielectric shells}

The form of the electrostatic interaction in Eq.~\eqref{eq:FES} was derived in Ref.~\onlinecite{bozic2013} for ion-permeable shells. Later work established that the theory can be extended to impermeable dielectric shells ($\varepsilon=\varepsilon_p$ for $r<R$, with the Poisson equation valid in the interior) in a first-order approximation~\cite{bozic2018,gnidovec2025}. In particular, the exact solution for a single such particle can be obtained from the solution of an ion-permeable particle through $g_l\rightarrow g_l^{\rm d}$, where
\begin{equation}
g_{l}^{\rm d}=\frac{\Cd(l,\kappa R)}{\kappa\varepsilon_w\varepsilon_0},
\label{eq:Cd}
\end{equation}
with
\begin{equation}
\Cd(l,x)=\Big[\frac{k_{l+1}(x)}{k_{l}(x)}+\frac{l}{x}\Big(\frac{\epsp}{\epsw}-1\Big)\Big]^{-1}.
\end{equation}
For an {ion-excluded} interior at $\epsp=\epsw$, Eq.~\eqref{eq:Cd} reduces to $\Cd=k_{l}/k_{l+1}$, the pure ion-exclusion response.

To obtain the pair interaction potential at first order in the dielectric contrast, one substitutes $\Cnull\!\to\!\Cd$;  writing $\xi_l=\Cd(l,x)/\Cnull(l,x)$, \rev{each single-particle response is replaced by its impermeable counterpart, $g_l\rightarrow g_l^d=\xi_l\,g_l$. Each interaction (off-diagonal) element acquires one factor for each of the particles, resulting in $\mathsf{G}_{(lm),(pm)}\rightarrow\mathsf{G}^\mathrm{d}_{(lm),(pm)} = \xi_l\,\xi_p\;\mathsf{G}_{(lm),(pm)}$, so that the coupling of two modes carries both their individual dielectric factors.} The same physical content---replacing an ion-impenetrable particle by an ion-penetrable charge distribution that reproduces its exterior potential---underlies the singular-charge mapping of \citet{everts2020}, where the analogue of \rev{the factor $\xi_l$} is a charge parameter $\Upsilon$ obtained by fitting the numerically computed surface potential. Our $\xi_l$ is the closed-form, mode-resolved counterpart, exact to first order in the dielectric contrast. At $l=0$ the factor reduces to $\xi_0 = x\,\mathrm{e}^{x}/[(1+x)\sinh x]$ with $x=\kappa R$, which is exactly the ratio between the DH potential of an ion-excluded sphere and the convolution result for an ion-penetrable one derived in the appendix of Ref.~\onlinecite{Trizac2002clay}.

The form of $\Gmat^\mathrm{d}$ can be used in place of $\Gmat$ wherever the interaction of impermeable dielectric shells is considered, and it captures the first-order dielectric response. Exact DH treatments of dielectric spheres are available~\cite{siryk2021,obolensky2021} and have recently been extended to many-body configurations~\cite{siryk2025a,siryk2025b}; they are, however, formulated at fixed charge. Here, we keep the first-order dielectric contrast deliberately, because it is what preserves the closed form of the CR step.

\subsection{Charge regulation in the surface density picture}

We now endow the spherical particles with a set $\mathcal{S}$ of ionizable species. Rather than index individual sites, we describe a species $j\in\mathcal{S}$ by its surface number density $n_{j}^{(i)}(\Om)$, a unit valence sign $\nu_{j}\in\{+1,-1\}$ (positive for bases, charged $+e$ when protonated; negative for acids, charged $-e$ when deprotonated), and a dissociation constant $\mathrm{p}K_j$. The instantaneous charge density is the Langmuir (Ninham--Parsegian) sum over the species~\cite{ninham1971,bozic2018,podgornik2018},
\begin{equation}
\sig(\Om)=e\sum_{j\in\mathcal{S}}\nu_{j}\,n_{j}(\Om)\,\eta_{j}(\Om),
\label{eq:CRnl}
\end{equation}
with
\begin{equation}
\eta_{j}(\Om)=\left[{1+10^{\,\nu_{j}(\mathrm{pH}-\mathrm{p}K_j)}\,\ee^{\,\nu_{j}\phi(\Om)}}\right]^{-1},
\end{equation}
where $\eta_{j}$ is the fraction of species $j$ in its charged state (protonated fraction for a base and deprotonated fraction for an acid); the local
potential induces a shift $\mathrm{pH}\to\mathrm{pH}+\phi(\Om)/\ln 10$.

In line with the linearization of the PB equation, which requires $|\phi|\lesssim1$, we also linearize each $\eta_{j}$ about $\phi=0$. Defining the bare charged fractions and site capacitances
\begin{eqnarray}
\vartheta_{j}\equiv\eta_{j}(0)&=&\left[1+10^{\,\nu_{j}(\mathrm{pH}-\mathrm{p}K_{j})}\right]^{-1},
\\
c_{j}\equiv\vartheta_{j}(1-\vartheta_{j})
&=&\frac{10^{\,\nu_{j}(\mathrm{pH}-\mathrm{p}K_{j})}}{\big[1+10^{\,\nu_{j}(\mathrm{pH}-\mathrm{p}K_{j})}\big]^{2}},
\label{eq:barefrac}
\end{eqnarray}
and noting $\partial_{\phi}\eta_{j}\big|_{0}=-\nu_{j}c_{j}$, we have $\eta_{j}\approx\vartheta_{j}-\nu_{j}c_{j}\phi$. We thus obtain a {local, linear} constitutive relationship,
\begin{equation}
\sigma(\Om)=\sigma_0(\Om)-\mathcal{C}(\Om)\,\phi(\Om),
\label{eq:closure}
\end{equation}
in which $\sigma_0(\Om)=e\sum_{j}\nu_{j}\,n_{j}(\Om)\,\vartheta_{j}$ is the bare (reservoir) charge density and the capacitance
\begin{equation}
    \mathcal{C}(\Om)=e\sum_{j}n_{j}(\Om)\,\vartheta_{j}(1-\vartheta_j)
    \label{eq:capacitance}
\end{equation}
is set by the (bare) local degree of ionization $\eta_j(0)$ of each species. Equation~\eqref{eq:closure} is the spatially resolved, multi-species generalization of the linearized (``constant-regulation'') boundary condition of planar CR theory~\cite{carnie1993,pericet2004,markovich2016}.

\subsection{Unified system and interaction energy}
\label{sec:join}

The total free energy of the charge-regulating system is $F=F_{\rm ES}+F_{\rm CR}$, where a CR term $F_\mathrm{CR}$ is added to the electrostatic free energy~\cite{bozic2018}. Expanding the exact site (mixing/binding) free energy to quadratic order about the bare state---the same linearization that produced the closure in Eq.~\eqref{eq:closure}---gives a {harmonic} penalty in which each charge mode is tethered to its {bare value} $\szero$ (i.e., fixed-charge value at $\phi=0$) with a stiffness set by the {inverse} capacitance,
\begin{equation}
F_{\rm CR}[\svec]=\frac{R^{2}}{2\beta e}\,(\svec-\szero)^{\dagger}\,\Kmat^{-1}\,(\svec-\szero).
\label{eq:FCR}
\end{equation}
Here, the {capacitance operator} $\Kmat$ is the multipole representation of the capacitance $\mathcal{C}(\Om)=\sum_{L,M}\mathcal{C}_{LM}Y_{LM}(\Omega)$,
\begin{eqnarray}
\Kmat_{(lm),(l'm')}&=&\oint \Ylmc(\Om)\,\mathcal{C}(\Om)\,Y_{l'm'}(\Om)\dO\\
\nonumber&=&\sum_{L,M}\mathcal{C}_{LM}\,\underbrace{\oint \Ylmc(\Omega)\,Y_{LM}(\Omega)\,Y_{l'm'}(\Omega)\dO}_{\text{Gaunt}} .
\label{eq:Kmat}
\end{eqnarray}
The Gaunt integral is nonzero only for
\begin{equation}
M=m-m',\quad |l-l'|\le L\le l+l',\quad l+l'+L\ \text{even},
\label{eq:selrule}
\end{equation}
so a capacitance multipole $\mathcal{C}_{LM}$ couples charge modes $(l',m')$ and
$(l,m'{+}M)$. \rev{The operator $\Kmat$ is positive semidefinite; whenever $\mathcal{C}(\Omega)$ is positive on some part of the surface, as it is for any continuous site density $n(\Omega)$, it is also positive definite and hence invertible (Appendix~\ref{app:A}). For point-like sites, $\Kmat$ can be singular; Eq.~\eqref{eq:FCR} is then understood on the range of $\Kmat$ and the results below, which involve $\Kmat$ but not its inverse, remain well defined.
}

The physical charges minimize the total free energy, and the variational solution $\delta F/\delta\svec\,|_{\sstar}=0$ gives the central result
\begin{equation}
\big[\Imat+\beta e\,\Kmat\,\Gmat(\rho,\omega_{1},\omega_{2})\big]\,\sstar=\szero,
\label{eq:master}
\end{equation}
a linear system for the regulated surface charge densities $\sstar(\rho,\omega_{1},\omega_{2})$ which now depend on separation and mutual orientation; $\Imat$ is the identity matrix. Equation~\eqref{eq:master} is the two-particle density-picture analogue of the site-indexed system in Ref.~\onlinecite{bozic2018}. Defining the \rev{dressed response operator including the CR mechanism} as
\begin{equation}
\Gmat_{\star}\equiv\Gmat\,(\Imat+\beta e\,\Kmat\Gmat)^{-1}
=(\Gmat^{-1}+\beta e\,\Kmat)^{-1},
\label{eq:Gstar}
\end{equation}
the total free energy evaluated at the minimum collapses to
$F_{\star}=F(\sstar)=\tfrac{R^{2}}{2}\,\szero{}^{\dagger}\Gmat_{\star}\szero$.
The interaction energy of the system is thus
\begin{equation}
V_{\rm int}(\rho,\omega_{1},\omega_{2})
=\frac{R^{2}}{2}\,\szero{}^{\dagger}
\big[\Gmat_{\star}(\rho,\omega_{1},\omega_{2})-\Gmat_{\star}(\rho\to\infty)\big]
\szero .
\label{eq:Vint}
\end{equation}
Because $\Gmat$ is fixed once we have set $\bm{\rho}\parallel\hat{z}$, orientations enter only through the intrinsic (body-frame) coefficients of each particle when rotated by $\omega_i$, affecting $n^{(i)}(\Omega)$ and hence $\szero$ and $\Kmat$. The product $\svec^{(0)\dagger}\Gmat_\star(\rho\to\infty)\,\svec^{(0)}$ is  the sum of two single-particle self-energies, and since rotating a particle rotates $\svec^{(0)}$ and $\Kmat$ together, the scalar is invariant to rotation. When $\Kmat\to0$ (no titratable capacitance), it follows that $\Gmat_{\star}\to\Gmat$ and $\sstar\to\szero$, and Eq.~\eqref{eq:Vint} reduces to the fixed-charge shell interaction of Ref.~\onlinecite{bozic2013}. The Gaunt mixing in $\Kmat$ reproduces the anomalous multipole expansion of the point-site theory~\cite{bozic2018} as a special case.

\subsection{Charge fluctuations}

Equation~\eqref{eq:master} determines the {most probable} charges---the mean-field value of the fluctuating ionisation degrees of freedom. The same degrees of freedom also {fluctuate} thermally, and when the fluctuations on the two particles are correlated they generate an additional interaction, the fluctuation (KS) force~\cite{kirkwood1952}. It is a distinct contribution to the free energy which follows from the {same} quadratic functional of  Eqs.~\eqref{eq:FCR}--\eqref{eq:Vint} by integrating the charges out rather than extremising over them.

To derive the KS interaction, we treat the multipole amplitudes $\svec$ as fluctuating fields with the Boltzmann weight $e^{-\beta F[\svec]}$, where $F[\svec]$ is the total free energy. Because $F$ is quadratic, the partition function is a Gaussian integral whose Hessian in the $(\svec^{\dagger},\svec)$ variables is
\begin{equation}
\mathcal{H}=R^{2}\Big(\Gmat+\tfrac{1}{\beta e}\Kmat^{-1}\Big)
=\frac{R^{2}}{\beta e}\,\Kmat^{-1}\big(\Imat+\beta e\,\Kmat\,\Gmat\big).
\label{eq:Hessian}
\end{equation}
Splitting off the saddle point $\sstar$ and setting
$\svec=\sstar+\delta\svec$,
\begin{eqnarray}
\nonumber Z&=&\int\!\mathcal{D}\svec\,\ee^{-\beta F[\svec]} \\
\nonumber &=&\ee^{-\beta F_{\star}}\!\int\!\mathcal{D}\delta\svec\,
\ee^{-\tfrac{\beta}{2}\delta\svec^{\dagger}\mathcal{H}\,\delta\svec} \\
&=&\ee^{-\beta F_{\star}}\,\big(\det\mathcal{H}\big)^{-1/2}\times\text{const},
\end{eqnarray}
so the free energy separates cleanly into the mean-field contribution
$F_{\star}$ (Sec.~\ref{sec:join}) and a fluctuation contribution
$\tfrac12 k_{\rm B}T\ln\det\mathcal{H}$. We thus obtain KS fluctuation interaction in closed form,
\begin{eqnarray}
\nonumber V_{\rm KS}(\rho,\omega_1,\omega_2)&=&\frac{\kB T}{2}\,\mathrm{Tr}\ln
\Big[\big(\Imat+\beta e\,\Kmat\,\Gmat(\rho,\omega_1,\omega_2)\big)\\
&&\times\big(\Imat+\beta e\,\Kmat\,\Gmat(\rho\to\infty)\big)^{-1}\Big].
\label{eq:KS}
\end{eqnarray}
Equation~\eqref{eq:KS} is the full anisotropic CR fluctuation interaction. It retains all multipole fluctuations and their cross-correlations, and depends on the mutual orientation of the particles through $\Kmat$. It is the two-body, density-picture form of the site-fluctuation determinant from Ref.~\onlinecite{bozic2021} and generalizes the solution of Ad\v{z}i\'c and Podgornik~\cite{adzic2014,adzic2015}, who obtained the KS interaction as the one-loop correction to the mean field for isotropic macroions. Its leading (monopole) term recovers the weak-coupling limit of Lund and J\"onsson~\cite{lund2005,lund2013}.

\subsection{Validity of the linearized approximation}
\label{sec:linear}

Derivation of the theoretical framework resorts to the linearized approximation twice: it solves the DH equation instead of the full non-linear PB system (Sec.~\ref{ssec:DH}) and it linearizes the CR closure [Eq.~\eqref{eq:closure}] when constructing the full free energy of the system (Sec.~\ref{sec:join}). Both are controlled by the same dimensionless parameter~\cite{deserno2002osmotic,andelman1995electrostatic,pericet2004}, the extremum of the reduced surface potential, which needs to be small: $\max_\Omega|\phi(\Omega)|\lesssim1$. The net charge $Q$ ($l=0$) provides the familiar lower bound, $\phi_0=\lB Q/[R\,(1+\kappa R)]\lesssim1$. For the CR closure, the error of linearising is $\sim\frac{1}{2}(1-2\vartheta_j)\phi$, meaning that the linear closure is best when the capacitance is the largest.  Where this parameter is not small, the established remedy is charge renormalization\cite{alexander1984charge,TrizacBocquetAubouy2002}, in which the nonlinear response of the ion cloud is absorbed into an effective charge fed back into a linear theory. Within PB theory a highly charged object behaves in the far field as a constant-potential particle at $\phi \simeq 4$, essentially independently of its shape\cite{TrizacBocquetAubouy2002}, and this prescription has been carried through in closed form for anisotropic particles\cite{Trizac2002clay}. Renormalization becomes quantitatively relevant above $\ell_BQ/R \gtrsim 5$\cite{brito2023effective}; for the protein models of Sec.~\ref{ssec:protein} this ratio is of order $3$, so the results there sit below the onset.

\section{Results}\label{sec:results}
In the following, we first explore the implications of the derived theory on simple cases of particles with the symmetry of a dipole and a linear quadrupole (Sec.~\ref{ssec:simple}). We compare the predictions of the theory with previous works in Sec.~\ref{ssec:benchmark} and apply it to proteins in Sec.~\ref{ssec:protein}.

\subsection{Charge regulation between anisotropic bodies with dipolar and quadrupolar symmetry}
\label{ssec:simple}

\begin{figure}[t]
\centering
\includegraphics{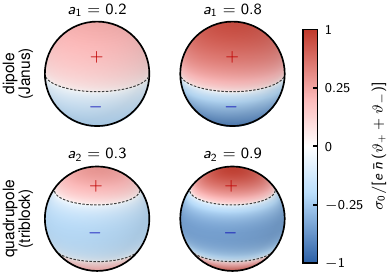}
\caption{\rev{Bare surface charge density $\sigma_0$ of the dipolar ($l=1$) and quadrupolar ($l=2$) models of Sec.~\ref{ssec:simple} at the pI, for a low and a high anisotropy contrast $a_l$. The color shows $\sigma_0$ in units of the mean density of ionized groups, $e\,\bar n(\vartheta_++\vartheta_-)$; at the pI this equals the local excess of basic over acidic sites, $(n_+-n_-)/(n_++n_-)=a_lP_l(\cos\theta)$, so that $\pm1$ corresponds to a region covered only by basic (acidic) sites. Dashed lines mark the nodal lines, $P_l(\cos\theta)=0$; the color scale is
compressed ($\propto|x|^{1/2}$) to keep the low-contrast particles visible. The high anisotropy contrast cases are the ones used in Fig.~\ref{fig:pi}.}
}
\label{fig:model}
\end{figure}

\begin{figure*}[t]
\centering
\includegraphics{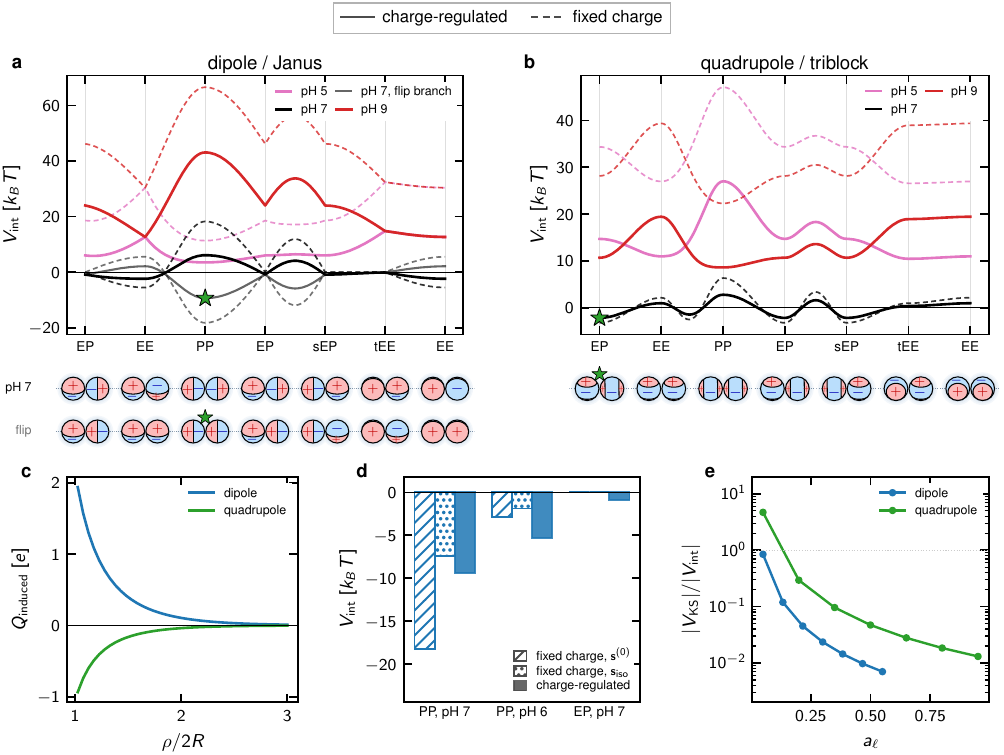}
\caption{CR effects in the interaction of two anisotropic particles with dipolar (Janus) and linear quadrupolar (triblock) symmetry. Conditions common to all panels unless stated otherwise: $R=2$~nm, $\bar{n}=0.7$~nm$^{-2}$, $\mathrm{p}K_{-}=6$ and $\mathrm{p}K_{+}=8$ (giving pI~$7$), $c_0=20$~mM ($\kappa R=0.93$), the pH is $7$, and ``contact'' means $\rho/2R=1.05$. The dipolar model uses $a_{1}=0.8$ and the quadrupolar one $a_{2}=0.9$, except where the anisotropy contrast is swept.
{\bf (a)}, {\bf (b)} Interaction along the one-parametric orientational pathway of Gnidovec \emph{et al.}~\cite{gnidovec2025} at contact.  The main orientations of the two particles relate the relative orientations of their equators (E) and poles (P); $\mathrm{s}$---staggered, $\mathrm{t}$---twisted. Star symbols mark the orientations used in other panels. \rev{The flip branch of the Janus particles follows the same pathway with the symmetry axis of the second particle reversed at every point, which exchanges its acidic and basic hemispheres; it is drawn at the pI only. Triblock particles are invariant under the flip.} \rev{Sketches below panels (a) and (b) show the pair orientations at the labeled configurations; for the Janus particles, the lower row shows the flip branch.}
{\bf (c)} {CR-induced monopole $Q_\mathrm{induced}$ carried by {one} particle as a function of particle-particle distance; it vanishes identically at fixed charge. \rev{For the dipole (PP, oppositely charged poles facing), the pair is antisymmetric under inversion about its midpoint, so that the second particle carries $-Q_{\rm induced}$ and the total charge of the pair is unchanged; for the quadrupole (EP), no such symmetry exists.}
}
{\bf (d)} Interaction energy of two dipolar particles at contact for bare fixed-charge conditions ($\svec^{(0)}$), conditions where the particles are titrated in isolation at a specific pH and their charge then held fixed [$\svec_\mathrm{iso}$, Eq.~\eqref{eq:siso}], and full CR conditions.
{\bf (e)} KS weight $|V_{\rm KS}|/|V_{\rm int}|$ at contact as a function of anisotropy contrast.
}
\label{fig:pi}
\end{figure*}

We first consider particles with simple symmetries---that of a dipole and a linear quadrupole, characteristic of Janus and triblock particles, respectively \rev{(Fig.~\ref{fig:model})}. We create minimal cases and set $\mathcal{S}=\{+,-\}$: one basic species ($\nu_{+}=+1$, density $n_{+}$, $\mathrm{p}K_{+}$) and one acidic species ($\nu_{-}=-1$, density $n_{-}$,
$\mathrm{p}K_{-}$). Then Eq.~\eqref{eq:CRnl} reads
\begin{equation}
\sig(\Om)=\frac{e\,n_{+}(\Om)}{1+\ee^{\,\phi+\ln10(\mathrm{pH}-\mathrm{p}K_{+})}}
-\frac{e\,n_{-}(\Om)}{1+\ee^{-\phi-\ln10(\mathrm{pH}-\mathrm{p}K_{-})}},
\end{equation}
and we have $\sig_0=e[n_{+}\vartheta_{+}-n_{-}\vartheta_{-}]$ and $\mathcal{C}=e[n_{+}c_{+}+n_{-}c_{-}]$. We choose site densities whose bare charge has the target symmetry with a minimal smooth model
\begin{equation}
    n_{\pm}(\Om)=\bar n\,[\,a_{0}\pm a_{l}P_l^0(\cos\theta)\,],
\end{equation}
where $l=1$ for dipole and $l=2$ for quadrupole particles. \rev{Here, $P_l^m$ are the associated Legendre polynomials, where we took into account that $m=0$ due to axial symmetry. As number densities, $n_\pm$ must be non-negative everywhere on the surface; this holds whenever $\lvert a_l\rvert\leqslant a_0$. We set $a_0=1$, so that $\bar{n}$ is the surface-averaged density of each species. The remaining parameter, the anisotropy contrast $a_l$, determines how the two species are distributed (Fig.~\ref{fig:model}), with $a_l=0$ reducing to a homogeneously charged sphere.} To keep the effects of CR visible, we set $\bar{n}=0.7$~nm$^{-2}$, even though the linear approximation breaks down here. \rev{We further set $\mathrm{p}K_-=6$ and $\mathrm{p}K_+=8$, which places the isoelectric point of the bare charge at $\mathrm{p}I=(\mathrm{p}K_++\mathrm{p}K_-)/2=7$. Unless stated otherwise, we also fix $R=2$~nm and $c_0=20$~mM ($\kappa R=0.93$).}

Figure~\ref{fig:pi} explores the behavior of the two models and lists the parameters used. In panels (a) and (b), we show the effect of orientational ordering through the one-parametric orientational pathways of \citet{gnidovec2025}, examining the main orientations anchored by the particles' equators (E) and poles (P) at different pH. At fixed-charge conditions (bare charge at a given pH), the interaction pathways match the previously studied ones~\cite{gnidovec2025}. In particular, at the isoelectric point ($\mathrm{pI}=7$ by construction), the Janus particles carry a pure odd-$l$ charge, \rev{so reversing (flipping) the axis of one particle is equivalent to $\sigma^{(2)}\rightarrow-\sigma^{(2)}$. As the charge distribution on one of the particles flips sign, we have at fixed charge $V_\mathrm{int}\rightarrow-V_\mathrm{int}$, and the two branches are exact mirror images (dashed black/grey lines in Fig.~\ref{fig:pi}a). CR breaks this (anti)symmetry: each particle induces a charge $Q_\mathrm{induced}$ on its partner (Fig.~\ref{fig:pi}c), and the interaction energy is quadratic in the induced charge and therefore does not change sign upon the flip and is always attractive. The two branches therefore split (full black/grey lines in Fig.~\ref{fig:pi}a), with their mean displaced towards attraction (e.g., by $\approx-1.6$~$\kB T$ at PP contact). This is a direct consequence of CR and has no counterpart in any linear fixed-charge model.}

\rev{At the pI, both particles are electroneutral in isolation, yet acquire a net charge $Q_\mathrm{induced}\neq0$ as they approach each other. Integrating Eq.~\eqref{eq:closure} over the surface of particle $i$ gives
\begin{equation}
    Q_\mathrm{induced}^{(i)}=-R^2\oint\dO\,\mathcal{C}^{(i)}(\Omega)\left[\phi_\star^{(i)}-\phi_\mathrm{iso}^{(i)}\right].
\end{equation}
This charge therefore vanishes at fixed charge and in configurations in which the partner's potential averages out over the particle's surface (e.g., two Janus particles in the EE configuration), but is non-zero for general orientations. The sign of the induced monopole is set by the charged species of the second particle nearer to the first one, and its magnitude decays with separation on the scale of the screening length (Fig.~\ref{fig:pi}c).}

\rev{The charge carried by an isolated charge-regulating particle, titrating in the screened potential of its own charges, is
\begin{equation}
\svec_{\rm iso}=\big[\Imat+\beta e\,\Kmat\,\Gmat_{\rm self}\big]^{-1}\szero ,
\label{eq:siso}
\end{equation}
where $\Gmat_{\rm self}$, i.e., $\Gmat$ with the off-diagonal blocks set to zero, accounts for these intra-particle interactions. Without them, the particle would carry its bare charge $\szero$, corresponding to ideal titration. The step $\szero\to\svec_{\rm iso}$ is a single-particle renormalization; only the further step $\svec_{\rm iso}\to\sstar$ is induced by the proximity of the second particle. Since the overall effect of CR depends on both the particle--particle orientation and the solution conditions, it is convenient to measure it against the bare fixed charge $\szero$. CR weakens an interaction if it reduces $\lvert V_\mathrm{int}\rvert$ and strengthens it if it increases $\lvert V_\mathrm{int}\rvert$. The single-particle renormalization $\szero\to\svec_{\rm iso}$ reduces $|\svec|$ and therefore moves $V_{\rm int}$ towards zero in all configurations of Fig.~\ref{fig:pi}, weakening repulsion and attraction alike. The proximity step $\svec_{\rm iso}\to\sstar$, by contrast, is sign-definite. The regulated charge minimizes $F$ at every $\rho$, and at $\rho\to\infty$ that minimizer is precisely $\svec_{\rm iso}$, so the reference term cancels and
\begin{equation}
V_{\rm int}(\sstar)-V_{\rm int}(\svec_{\rm iso})
= F(\sstar)-F(\svec_{\rm iso})\;\leqslant\;0
\label{eq:proxbound}
\end{equation}
at every separation and mutual orientation. The genuinely proximity-induced part of CR can therefore only make the interaction more attractive.
}

\rev{
The decomposition of contributions in Fig.~\ref{fig:pi}d demonstrates that the two steps of Eqs.~\eqref{eq:siso} and~\eqref{eq:proxbound} indeed act differently. The single-particle step $\szero\rightarrow\svec_\mathrm{iso}$ reduces the charges and, in all configurations of Fig.~\ref{fig:pi}, moves $V_\mathrm{int}$ towards zero, whereas the proximity step $\svec\rightarrow\sstar$ can only lower $V_\mathrm{int}$. A repulsion is therefore weakened (all curves at pH 5 and 9 in Fig.~\ref{fig:pi}a,b), whereas for an attraction the two steps compete (Fig.~\ref{fig:pi}d). For dipolar particles at contact in the PP configuration at pH 7, the single-particle step dominates and the attraction is weakened by a factor of $\approx 2$; at pH 6, the proximity step dominates and the attraction is strengthened. In the EP configuration at pH 7---the isoelectric point---both fixed-charge references give zero, and the proximity step alone creates an attraction of $\approx-1$~$\kB T$. Depending on the system, CR can thus either weaken or strengthen the interaction between two particles, and even lead to the appearance of an attraction between two otherwise non-interacting particles.
}

Beyond the mean-field CR interaction, the KS interaction is a monopole charge-fluctuation effect and is set by how many sites titrate and how strongly, not by how they are arranged (when $\mathcal{C}$ is uniform). For the two particle models under consideration here, the KS interaction $V_{\rm KS}$ turns out to provide a fraction of the total CR interaction under most values of the anisotropy contrast (Fig.~\ref{fig:pi}e), falling as \rev{the anisotropy contrast} $a_l$ increases. The divergence as $a_l\to0$ is physical---for a perfectly isotropic regulator at its isoelectric point the mean field vanishes identically and the fluctuation term is the entire interaction, recovering the original KS problem~\cite{kirkwood1952} as a limit of the present theory. \rev{For a protein with discrete sites, the anisotropy contrast lies within the range of Fig.~\ref{fig:pi}e: for the two protein models of Sec.~\ref{ssec:protein} at pH 7, we obtain $a_1=0.17$ and $a_2=0.34$ for lysozyme and $a_1=0.39$ and $a_2=0.74$ for $\alpha$-chymotrypsinogen A~\cite{bozic2017ph,felder2007server}. However, the KS weight is then set mainly by the capacitance at the given pH and by the net charge rather than by the protein anisotropy. For both proteins at contact and $c_0=20$~mM, the KS term changes the orientationally averaged interaction by less than $0.01\%$ at pH 7 and by at most $\sim 1\%$ at pH 4, where the carboxyls titrate; it is therefore negligible for the proteins considered here.}

\subsection{Validation against three benchmarks}
\label{ssec:benchmark}

\begin{figure*}[t]
\centering
\includegraphics{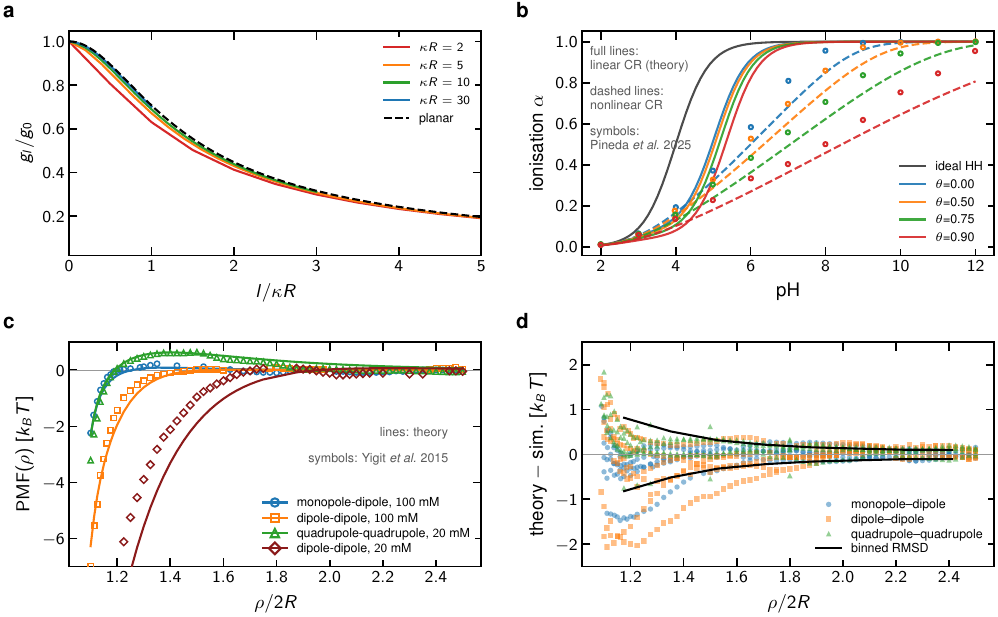}
\caption{Benchmarking the theory. {\bf (a)} Normalized self-response of differently-sized charge-regulating patches as given by their wavenumber. Dashed line shows the prediction of Ref.~\onlinecite{boon2011} for a flat surface: a lateral mode of wavenumber $k$ is screened over its own size and follows $[1+(k/\kappa)^{2}]^{-1/2}$. On a sphere, degree $l$ is a feature of wavenumber $k\simeq l/R$, and $g_{l}/g_{0}$ collapses onto that planar form as the mode stops
seeing the curvature. {\bf (b)} Titration of a single patchy CR sphere against grand-reaction MC~\cite{pineda2025} (symbols) for polar caps of decreasing coverage (increasing $\theta$). Full lines show the {linearised} closure used throughout this work, dashed lines the self-consistent nonlinear single-particle titration. {\bf (c)} Comparison of the fixed-charge theoretical predictions against the explicit-salt CPPM simulations of Ref.~\onlinecite{yigit2015} in the form of orientationally averaged PMFs that include electrostatic interaction and the neutral-sphere Mie baseline. Four out of $16$ cases are shown: the best-agreeing pair (lowest RMSD) in each of the three interaction families---monopole--dipole, dipole--dipole, quadrupole--quadrupole---together with the single worst-agreeing low-salt case (dipole--dipole at $20$~mM salt concentration). {\bf (d)} Same system as in panel (c), showing the residual in PMFs obtained by theory and simulation for all $16$ cases studied in Ref.~\onlinecite{yigit2015}. Full line shows the distance-binned per-pair RMSD.
}
\label{fig:bench}
\end{figure*}

We benchmark the developed theory against three published works~\cite{boon2011,yigit2015,pineda2025} that together help elucidate three independent aspects of the theory: the single-particle mode response $g_{l}$ that forms the diagonal of $\Gmat$, the off-diagonal block of $\Gmat$ that carries the anisotropy and the orientational average, and the closure of Eq.~\eqref{eq:closure} that connects the potential with charge. The benchmarks are gathered in Fig.~\ref{fig:bench}; to the best of our knowledge, there is no published work which would allow for a direct comparison with the developed theory as a whole.

\citet{boon2011} solved the CR problem on a patchy {plane} and found that a patch charges up to a higher density the smaller it is. Since a lateral mode of wavenumber $k$ is screened over its own size, $\kappa_{k}=\sqrt{\kappa^{2}+k^{2}}$, rather than over the Debye length, a fine feature has a smaller self-response. Our theory must contain that result in the flat limit ($R\to\infty$): identifying a spherical harmonic of degree $l$ with a lateral wavenumber $k\simeq l/R$, the mode response $g_{l}/g_{0}$ collapses onto the planar form of Ref.~\onlinecite{boon2011} as the mode stops seeing the curvature (Fig.~\ref{fig:bench}a). This check---made on a single particle---establishes that the screened multipole response carries the correct dependence on feature size, not merely the correct monopole. The two-surface continuation of that work\cite{boon2012electrostatics} is the closest existing analogue of the pair result derived here, and it also marks a boundary of the present treatment. There, the crossover from long-range repulsion to short-range attraction appears only well inside the nonlinear regime, and is reported to be absent from the linearized one. Our closure therefore reproduces the mechanism and its ordering with feature size, but should not be expected to locate the crossover quantitatively at comparable surface potentials. This also underlines why the multipole structure cannot be dropped at higher salt: unlike in vacuum, where the far field of any bounded charge distribution becomes isotropic, in an electrolyte the anisotropy survives at all distances and in fact becomes {more} pronounced as $\kappa$ grows~\cite{Trizac2002clay,everts2020}.

The grand-reaction MC simulations of \citet{pineda2025} titrate $10$ acidic sites on a polar cap of tunable coverage. Our linearized theory qualitatively tracks them on a single particle across the whole patchiness series with no fitted parameter (Fig.~\ref{fig:bench}b), reproducing the departure from ideal Henderson--Hasselbalch form and its growth as the cap shrinks from the geometry alone. The {linearised} form we use, which is what makes the theory analytic, carries a mean RMSD of $0.2$ against $0.06$ for the nonlinear form (self-consistent single-particle titration; not used anywhere in the two-body theory), degrading systematically with patchiness $\theta$ ($0.12$ at $\theta=0$ to $0.31$ at $\theta=0.9$). This shows that the linearisation is a real approximation, performing the worst exactly where the anisotropy is strongest.

Yigit \emph{et al.}~\cite{yigit2015} studied the interaction of charged patchy protein models (CPPMs)---impermeable spherical globules with specific localized surface charges (patches) matching real protein multipole moments---in the fixed-charge limit using explicit salt simulations. They considered the interactions of particles with low-symmetry charge distributions---monopole--dipole, dipole--dipole, and quadrupole--quadrupole---at two salt concentrations. As Fig.~\ref{fig:bench}c demonstrates, our theory (also evaluated at fixed charge) captures the simulation-determined PMFs very well, with the exception of a few cases at weak screening. Full comparison of the difference between the predictions of our theory and the $16$ pairwise interaction curves studied through simulations reveals a mean per-pair RMSD of $0.39\,\kB T$ (Fig.~\ref{fig:bench}d), where the deviation is confined to the deepest contact wells.

The three benchmarks together thus outline where the developed theory fails. In Ref.~\onlinecite{boon2011}, nonlinear PB charges up fine features more than the linearised estimate, and our theory is likely to follow suit (as already indicated in Sec.~\ref{sec:linear}). Similarly, as the titration sites on a particle become denser~\cite{pineda2025}, the nonlinear closure outperforms the linearized one, even though both follow the correct trend. And the agreement with the (fixed-charge) CPPM model degrades at weak screening and in the deepest contact wells, which is again where linearization of the PB equation is too strong of an approximation. All three benchmarks show the same---linearized response loses accuracy when local potentials stop being small. The natural next step that would help define the limits of our theory would be benchmarking it against constant-pH or grand-reaction simulation of two patchy regulating particles.

\subsection{Application to proteins}
\label{ssec:protein}

\begin{figure}[t]
\centering
\includegraphics{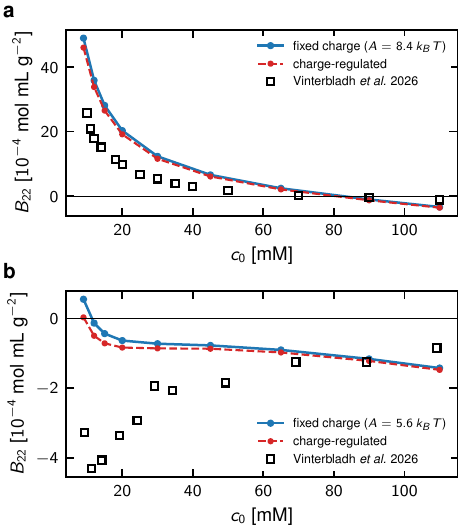}
\caption{Theoretically predicted $\Bt$ compared with SLS experiments~\cite{vinterbladh2026intermolecular}. The theory contains {no adjustable electrostatic parameter}; the single fit per protein is the non-electrostatic short-range attraction constant $A$. The comparison is shown for {\bf (a)} lysozyme ($R_{95}=2.13$~nm) and {\bf (b)} $\alpha$-chymotrypsinogen A ($R_{95}=2.38$~nm).
}
\label{fig:b22}
\end{figure}

Lastly, we show how the theory can be applied to proteins. Specifically, we focus on lysozyme and $\alpha$-chymotrypsinogen A, for which recent measurements of $\Bt$ are available~\cite{vinterbladh2026intermolecular}. We use a simple model with interior permittivity $\varepsilon_p=4$ where we map the ionizable amino acid residues onto an ion-impermeable sphere from the protein crystal structure (PDB: 4LZT for lysozyme and PDB: 7KTZ for $\alpha$-chymotrypsinogen A) and use their bulk $\pKa$ values. Each ionizable group $k$ is placed at its atomic centroid $\bm{r}_k$ and positioned on a sphere whose radius is the ``circumscribing'' (95th-percentile charge-group) radius $R_{95}$ using multipole weights $(\lvert\bm{r}_k\rvert/R_{95})^l$; this is done so that the two-body multipole expansion is valid at contact. The orientationally averaged protein--protein interaction $w(\rho)$ is determined as
\begin{equation}
    w(\rho)=w_\mathrm{EL}(\rho)+w_\mathrm{vdW}(\rho),
    \label{eq:w}
\end{equation}
where
\begin{equation}
    w_\mathrm{EL}(\rho)=-\kB T\,\ln\left\langle e^{-\beta V_\mathrm{int}(\rho,\Omega)}\right\rangle_{\Omega}
    \label{eq:wel}
\end{equation}
and $\Omega$ is the five-dimensional space covering all possible mutual orientations of the two proteins. The non-electrostatic part is given by the van der Waals interaction
\begin{equation}
    w_\mathrm{vdW}(\rho)=-\frac{A}{6}\left[\frac{2R^2}{\rho^2-4R^2}+\frac{2R^2}{\rho^2}+\ln\left(1-\frac{4R^2}{\rho^2}\right)\right],
    \label{eq:vdw}
\end{equation}
limited to distances above $\rho\geqslant2R_{95}+0.35\,\mathrm{nm}$; here and in Eq.~\eqref{eq:wel}, $R\equiv R_{95}$. Equation~\eqref{eq:vdw} is the source of the single fit parameter for each protein, the short-range (Hamaker) attraction parameter $A$. The electrostatic model of the protein is drawn directly from the crystal structure and contains no fit parameters. Against the static-light-scattering data of \citet{vinterbladh2026intermolecular} at pH~7 (Fig.~\ref{fig:b22}), lysozyme reproduces the strong low-salt repulsion screening toward the hard-sphere value, while $\alpha$-chymotrypsinogen A is attractive ($\Bt<0$) despite a net charge of $+4e$. Even though CR does not lead to a large effect in this case, the comparison in Fig.~\ref{fig:b22} establishes that the theory contains sufficient detail to reproduce the general trends of protein--protein interaction.

\begin{figure}[t]
\centering
\includegraphics{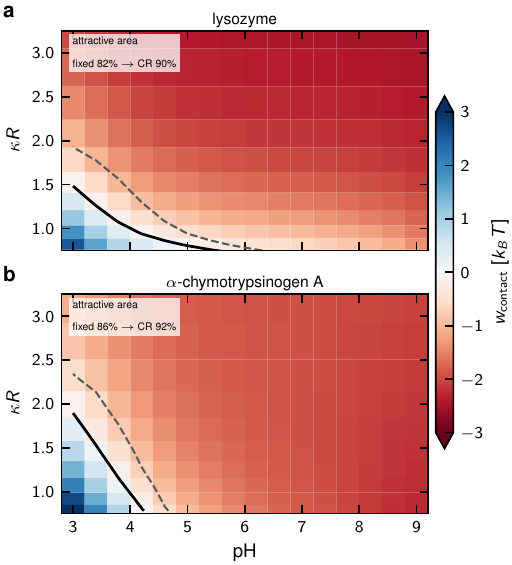}
\caption{CR expands the attractive basin of protein--protein interaction in the $(\mathrm{pH},\kappa R)$ space. Heatmaps show the orientationally averaged interaction of Eq.~\eqref{eq:w}, evaluated at contact ($\rho_\mathrm{contact}=2R_{95}+0.35\,\mathrm{nm}$), of {\bf (a)} lysozyme and {\bf (b)} $\alpha$-chymotrypsinogen A. Solid line is the zero contour of the charge-regulated model, whereas dashed line shows the zero contour of the fixed-charge model. CR enlarges the attractive basin from $82\%$ to $90\%$ in (a) and from $86\%$ to $92\%$ in (b). The Hamaker $A$ is inherited from the $\Bt$ fit of Fig.~\ref{fig:b22}, and the boundary depends on it.}
\label{fig:phase}
\end{figure}

Establishing the simple protein models allows us to examine what CR brings to the protein--protein interaction. In Fig.~\ref{fig:phase}, we show the aggregation heatmap of the two proteins in the $(\mathrm{pH},\kappa R)$ space. Lysozyme (pI $\simeq 11$) is repulsive at low salt and low pH; $\alpha$-chymotrypsinogen A (pI $\simeq 9.8$) is attractive over most of the plane. In both, CR {enlarges} the attractive basin---from $82\%$ to $90\%$ of the mapped area for lysozyme and from $86\%$ to $92\%$ for chymotrypsinogen---with the shift concentrated where the carboxyls titrate, i.e. where the site capacitance is largest. {The direction is not accidental, and it follows from Eq.~\eqref{eq:proxbound}: relative to the bare charge, the single-particle step weakens the electrostatic repulsion that sets the zero contour, and the proximity step can only lower $V_{\rm int}$, at every separation and orientation. Both therefore enlarge the attractive basin.} We also note that the boundaries depend on the fitted parameter $A$ from Fig.~\ref{fig:b22}, and Fig.~\ref{fig:phase} should thus be seen as a map of a mechanism, not a quantitative aggregation map of these two proteins.

\section{Discussion}
\label{sec:discussion}
We derived a theoretical framework which unifies three components that have previously been treated separately---anisotropic DH electrostatics, multi-species CR mechanism, and charge fluctuations. Equation~\eqref{eq:master} is a single linear system whose fixed-charge limit is the classical two-shell interaction and whose CR component is a Gaunt-structured capacitance operator; the compact interaction energy of Eq.~\eqref{eq:Vint} and the KS
determinant Eq.~\eqref{eq:KS} follow without further approximations. The framework provides a closed-form theoretical description of the pairwise electrostatic interaction between two spherical particles that is broadly applicable to colloids and proteins alike, making it well suited to scanning formulation space (pH, salt) for proteins and patchy colloids, and to supplying analytically transparent input to coarse-grained models.

Applying the framework to simple model particles with dipolar and linear quadrupolar symmetry of ionizable sites (Fig.~\ref{fig:pi}), we observed that for like-charged, monopole-dominated pairs CR weakens the repulsion, bounded between the constant-charge and constant-potential limits as expected from planar theory. For anisotropic and mixed-sign pairs the same response can deepen an existing attraction or create one where fixed charge gives none, and between electroneutral bodies it induces a monopole---a concrete, falsifiable prediction for the isoelectric regime that no {linear} fixed-charge model contains. This effect is consistent with CR-driven clustering of neutral Janus particles seen in many-body simulations~\cite{yuan2022,curk2021}. In particular, dilute suspensions of overall charge-neutral Janus particles were found to undergo a conformational transition from open strings and bundles to compact clusters at pH~$\sim\mathrm{p}K$ under CR, {absent} under the constant-charge condition~\cite{yuan2022}. It was also shown that the transition is driven by electrostatics and that CR is necessary for this transition---anisotropic charge distribution alone is insufficient. Our theory provides an analytical pair-level counterpart to the many-body simulations. This extends to proteins, where the theory can provide a qualitative description of the protein--protein interaction despite its simplicity (Fig.~\ref{fig:b22}), and in doing so {shows the proximity-induced part of the response acting in the direction guaranteed by Eq.~\eqref{eq:proxbound}, enlarging the attractive basin in the $(\mathrm{pH},\kappa R)$ space (Fig.~\ref{fig:phase})}.

The limitation of the theory's quantitative reach is primarily bound to the linearization approximation of both the electrostatics (DH equation) and the CR closure. These are required to make the theory closed-form, but the results overestimate the interaction energy at conditions of low salt and high charge. To our knowledge, no simulation results for a pair of anisotropically charged, charge-regulating particles exist which would provide a natural assessment of the validity of the theory. Instead, we compared its prediction to three different sets of results (Fig.~\ref{fig:bench}), which in particular allowed us to show the limits of the CR closure on single-particle cases. A natural extension is to compose the present theory with charge renormalization. Because renormalization is a single-particle construction, it enters only through the inputs $\mathbf{s}^{(0)}$, $\mathsf{K}$ and $\kappa$, leaving Eqs.~\eqref{eq:master}--\eqref{eq:KS} untouched. \citet{Trizac2002clay} built a closed-form screened pair potential for anisotropic platelets carrying a saturated effective charge and showed that anisotropy and counter-ion condensation together produce phase behavior that neither yields alone. Their construction, however, rescales the overall charge while holding the anisotropy function fixed, and they note explicitly that in the saturation regime the shape of that function is itself a signature of the effective charge
distribution and is a priori unknown. That is precisely the mode-dependence that a multipole formulation supplies. Since $g_l$ falls with both $l$ and $\kappa R$, and since a mode of degree $l$ is screened over $\kappa_l = \sqrt{\kappa^2 + (l/R)^2}$ rather than over the Debye length (Fig.~\ref{fig:bench}a), \rev{fine features saturate later and are renormalized less. The renormalization operator acting on $\svec$ is thus mode-resolved and---provided that mode couplings of the kind found for Janus spheres~\cite{boon2010jpcm,boon2012electrostatics} are retained---non-diagonal in the multipole basis, just like the capacitance operator $\Kmat$.} This combination remains an open problem~\cite{everts2020,boon2012electrostatics}: renormalization is a
far-field construction and would not be expected to repair the contact wells of
Fig.~\ref{fig:bench}d, and the appropriate prescription at infinite dilution differs from the cell-model schemes assessed in Ref.~\onlinecite{brito2023effective}.

\begin{acknowledgments}
This work is dedicated to the memory of Rudi Podgornik. I thank Jeffrey Everts for valuable comments on the manuscript. The work was supported by the Slovenian Research and Innovation Agency (research project No.\ J1-60002 and research core funding No.\ P1-0055).
\end{acknowledgments}

\section*{Data availability}
The data {that support the findings of this study are} available from the author upon reasonable request.

\section*{Author declarations}
The authors have no conflicts to disclose.

\appendix

\section{Positivity of the capacitance operator}
\label{app:A}

\rev{
We have $\Kmat$ as the representation of the capacitance density [Eq.~\eqref{eq:capacitance}] on the spherical harmonics with $l\leqslant l_\mathrm{max}$ used in numerical calculations; for two particles, $\Kmat$ is block diagonal (one block per particle). For any set of multipole coefficients $\bm{x}$, its corresponding surface density is $f_{x}(\Omega)=\sum_{lm}x_{lm}Y_{lm}$. It follows that
\begin{equation}
    \bm{x}^\dagger\Kmat\bm{x}=\oint\dO\,\mathcal{C}(\Omega)\lvert f_x(\Omega)\rvert^2\geqslant0,
    \label{eq:xKx}
\end{equation}
since $\mathcal{C}(\Omega)\geqslant0$: each species contributes $en_j\vartheta_j(1-\vartheta_j)$, the variance of the charge state of its sites. Therefore, $\Kmat$ is positive semidefinite. The integral in Eq.~\eqref{eq:xKx} vanishes only if $f_x$ is zero wherever $\mathcal{C}>0$. If $\mathcal{C}$ is positive on some patch of the surface of the sphere, as for any continuous site density, $f_x$ must then vanish everywhere, because a finite sum of spherical harmonics cannot vanish on a patch of the sphere without vanishing identically. Hence, $\bm{x}=0$ and $\Kmat$ is positive definite. Furthermore, if $\mathcal{C}$ is bounded from below by $\mathcal{C}_\mathrm{min}>0$, Parseval's theorem gives $\bm{x}^\dagger\Kmat\bm{x}\geqslant\mathcal{C}_\mathrm{min}\, \bm{x}^\dagger\bm{x}$. This is the case for the models of Sec.~\ref{ssec:simple}, for which $\mathcal{C}(\Omega)\geqslant2e\,\bar{n}\,\min(c_+,c_-)$ whenever $\lvert a_l\rvert\leqslant 1$.
}

\rev{
For $N_s$ point-like sites, by contrast, $\Kmat$ is a sum of $N_s$ terms (one per site) and is singular when $N_s<(l_\mathrm{max}+1)^2$, because patterns that vanish at every site cannot be titrated and stay at their bare values. If the sites are instead smoothed into a density truncated at $l_\mathrm{max}$, $\Kmat$ remains positive semidefinite as long as that density is non-negative.
}

\rev{
The central results of the manuscript do not require $\Kmat$ to be invertible. $\Gmat$ is positive definite because $\svec^\dagger\Gmat\svec$ is the DH electrostatic energy of a charge distribution $\svec$, which is positive for any $\svec\neq0$. Since $\Kmat\Gmat$ has the same eigenvalues as $\Gmat^{1/2}\Kmat\Gmat^{1/2}$, which are real and non-negative by Eq.~\eqref{eq:xKx}, all eigenvalues of $1+\beta e\Kmat\Gmat$ are real and not smaller than one. Equations~\eqref{eq:master}--\eqref{eq:Vint} and~\eqref{eq:KS} are therefore well-defined for any positive semidefinite $\Kmat$. Finally, $\det(1+\beta e\Kmat\Gmat)=\det(1+\beta e\Kmat^{1/2}\Gmat\Kmat^{1/2})$, where the matrix on the right is positive definite and its diagonal blocks do not depend on the distance $\rho$. Switching on the off-diagonal blocks that couple the two particles can only lower the determinant by virtue of Fischer's inequality, so that
\begin{equation}
    \det\left[1+\beta e\Kmat\Gmat(\rho)\right]\leqslant\det\left[1+\beta e\Kmat\Gmat(\rho\to\infty)\right],
\end{equation}
i.e., the KS interaction is always attractive.
}

\bibliography{bibliography}

\end{document}